# Silent Tracker: In-band Beam Management for Soft Handover for mm-Wave Networks

Santosh Ganji, Tzu-Hsiang Lin, Jaewon Kim, P. R. Kumar

## ABSTRACT

In mm-wave networks, cell sizes are small due to high path and penetration losses. Mobiles need to frequently switch softly from one cell to another to preserve network connections and context. Each soft handover involves the mobile performing directional neighbor cell search, tracking cell beams, completing cell access request, and finally, context switching. The mobile must independently discover cell beams, derive timing information, and maintain beam alignment throughout the process to avoid packet loss and hard handover. We propose Silent Tracker which enables a mobile to reliably manage handover events by maintaining an aligned beam until the successful handover completion. It is an entirely in-band beam mechanism that does not need any side information. Experimental evaluations show that Silent tracker maintains the mobile's receive beam aligned to the potential target base station's transmit beam till the successful conclusion of handover in three mobility scenarios: human walk, device rotation, and 20 mph vehicular speed.

## 1 INTRODUCTION

Unlike previous generation omnidirectional cellular networks, the coverage area of the mm-wave cell is small [1] due to high path and penetration losses. The cell deployment density is accordingly higher, and a walking user will need to switch through several cells [8]. In directional 5G mm-wave networks, initial beam search can take up to 1.28 seconds. which disrupts most multi-media applications such as video calling, VR and online gaming. Therefore, mm-wave mobile devices need to jump-start this process by early discovery of a neighbor cell beam, and adapt its beams during transition to smoothly perform handover.

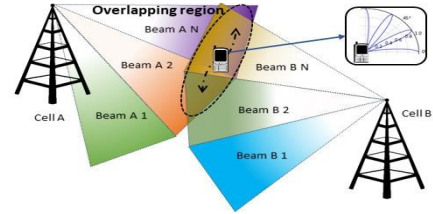

**Figure 1: mm-Wave mobile at the cell edge**

Such handovers need to be soft handovers that preserve network connections and context, since in the case of hard handover, the existing network connections are terminated and the mobile will need to connect to the new cell performing all the initial access procedures with no context and as if it were a newly joined user. This involves neighbor cell discovery, tracking both serving and neighbor cell beams, and random access at the physical layer. Also, coordination between all the layers in the full network stack is needed to perform authentication, network connection transfer, and context switching. This paper addresses the physical layer challenge of mobile side beam management during handover, that is critical for all subsequent upper layer connection transfer procedures.

A mobile must at all times strive to point its beam towards the base station for the much preferred line-of-sight communication that results in least path loss. While the mobile is moving within a cell, both mobile and base station therefore continually adapt their beams to maintain a high degree of alignment through exchange of control plane information [2]. This paper addresses the additional beam management needed for a mobile located at cell edge to perform a smooth handover to a neighboring base station.

## 2 CHALLENGES

Consider a mobile that is located at the edge of cell, say Cell A as shown in Fig. 1, at its boundary with Cell B. We focus on this transition regime. To avoid a hard handover, the mobile while staying connected to Cell A needs to perform initial cell access procedures that are time consuming to transit to a neighboring. The mobile must first perform directional cell search to find a neighboring cell, in this case Cell B. Then, it needs to continue to ḷtrack" a beam of Cell B, i.e., align one its beams with one of Cell B's beams, while moving in the overlapped coverage region. This needs to be done even while it is yet to get an access grant to Cell B.

The mobile must continue to maintain beam alignment with the original serving cell, i.e., Cell A, during the entire handover process to accomplish a soft handover. The mobile must therefore utilize its radio resources for measurements efficiently to accomplish beam alignment with both Cells A and B to avoid service interruption and hard handover. Within the limited measurement schedules available for serving Cell A and the unknown schedules of Cell B, the mobile must adapt two of its receive beams, one to preserve data connection to Cell A, and other for context transfer to Cell B.

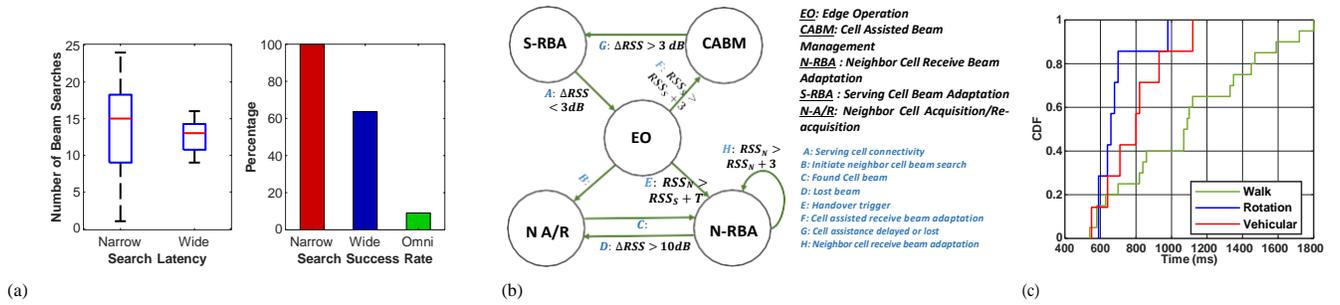

**Figure 2: (a) Mobility experiments at cell edge for Human walk. (b) Silent Tracker State Machine. (c) Silent Tracker evaluation**

We address this problem of beam management for soft handover. It needs to be done with minimal resource usage, while only utilizing the information available in the radio domain. Based on several observations made from extensive experiments under device and user mobility, we propose Silent Tracker, a minimalistic protocol for managing the mobile side beam, which avoids a hard handover.

Reactive handover mechanisms employed in omnidirectional cellular technologies are not viable in the mm-wave band, as directional search under mobility is highly delay prone. After searching for and discovering a neighbor cell's beam using directional receive radio beams, the mobile obtains timing synchronization, and transmits an uplink preamble, anticipating a response. The mobile needs to maintain a high degree of alignment of one of its beams to the newly found cell's beam, throughout the handover procedure. As the mobile is yet to establish a connection with the newly found cell, there is no assistance from that cell to adapt its beams, and the mobile must completely rely on "receive-beam adaptation".

Silent Tracker only requires Received Signal Strength (RSS) at the mobile, readily available in-band. It thereby differs from most of the existing works in the literature addressing the connected state beam alignment problem in directional communication networks. They typically rely on network-level coordination among the cells, exchanging measurement information such as angle of arrival acquired inband [7, 9] or out of band information from sensors [3] or side channels [10]. Motion prediction based handover management approaches [5, 6, 11] need location and direction of motion.

At cell edge, communication is unreliable with the serving cell (Cell A above), and not possible with neighboring cell (Cell B) before transition. Network level coordination to perform predictive handover is therefore infeasible or hard to accomplish in network deployments. It is necessary to develop protocols that only employ information that is acquired in-band, and to develop beam management protocols which are less resource-intensive, and with complete control at the mobile. This is what Silent Tracker provides.

## 3 SILENT TRACKER

Silent Tracker is, to our knowledge, the first in-band mechanism that maintains a beams to the serving cell while (silently) tracking a beam from a neighboring cell, thereby facilitating soft handover. It is based on extensive experiments made on a typical cell edge scenario with one mobile node and 3 nodes operating as base stations, created on our 60 GHz testbed [4]. Experimental observations have been made for the cases of a user walking with speed v= 1.4 m/s at the cell edge, 10 m from the base station, Device rotation ($\omega$ = 120 deg/s), and Vehicular motion (v= 20 mph). We performed several experiments to quantify latency and beam search success rate under user mobility at cell edge. We repeated the experiments employing different beamwidth (20°, 60°) codebooks and omni-directional beams at the mobile. Although search under mobility is highly delay prone, narrow beams have a significantly higher success rate than using an omnidirectional/single antenna at the mobile. The Human Walk results are presented in Fig. 2a.

The Silent Tracker protocol is shown in Fig. 2b. If it has not already discovered a neighboring cell's transmit beam, Silent Tracker does a search for one. As soon as such "initial search" discovers one or more beams of a neighboring cell (Cell B above), Silent Tracker begins the adaptation to its mobility of the mobile's receive beam for the neighboring cell. This is done by the following mobile side adaptation shown in Fig. 2b: Switch to one of the directionally adjacent receive beams when the RSS from the neighboring cell's base station drops by 3dB. All the while, Silent Tracker is also continually adapting the mobile's receive beam and the transmit beam from the serving cell (Cell A above), as detailed in the BeamSurfer Protocol [2]: (i) Mobile side adjustment: switching to one of the directionally adjacent receive beams for the serving cell when its RSS drops by 3dB, and (ii) Base station adjustment: switching to one of the directionally adjacent transmit beams when (i) no longer suffices. The latter requires communication from the mobile to the serving cell base station. Both adjustments only require knowledge of RSS, as detailed in the protocol in Fig. 2b. If the mobile is at cell edge, then at some point, it will not be able to communicate to the serving cell's (Cell A's) base station and adaptation (ii) is not possible, at which time the link to the serving cell is disrupted. At that point, Silent Tracker switches its serving cell to the neighboring cell (Cell B), and initiates a random access protocol to the neighboring cell's base station.

The evaluation of Silent Tracker at the edge of two cells in Fig. 2c shows that it is successful at adapting its receive beam for all three mobility scenarios considered. It successfully tracks the beams for the cases of a slowing moving user, as well as the quick transition cases of mobile device rotation and mobile movement at vehicular speed.